\documentclass[12pt,a4paper]{article}

\pdfoutput=1

\usepackage{ifthen,placeins} 
\usepackage{hyperref}
\usepackage{cite}
\usepackage{dcolumn}

\newcolumntype{d}[1]{D{.}{\cdot}{#1}}

\newboolean{pdflatex}
\setboolean{pdflatex}{true} 

\newboolean{articletitles}
\setboolean{articletitles}{true} 

\newboolean{uprightparticles}
\setboolean{uprightparticles}{false} 

\newboolean{inbibliography}
\setboolean{inbibliography}{false} 

\usepackage[top=1in, bottom=1.25in, left=1in, right=1in]{geometry}

\usepackage{comment}
\usepackage{microtype}
\usepackage{lineno}  
\usepackage{xspace} 
\usepackage{caption} 

\usepackage{graphicx}  
\usepackage{color}
\usepackage{colortbl}
\graphicspath{{./figs/}} 

\usepackage{amsmath} 
\usepackage{amssymb}
\usepackage{amsfonts}
\usepackage{upgreek} 

\usepackage{longtable} 
\usepackage{booktabs}

\newcommand*\patchAmsMathEnvironmentForLineno[1]{%
\expandafter\let\csname old#1\expandafter\endcsname\csname #1\endcsname
\expandafter\let\csname oldend#1\expandafter\endcsname\csname
end#1\endcsname
 \renewenvironment{#1}%
   {\linenomath\csname old#1\endcsname}%
   {\csname oldend#1\endcsname\endlinenomath}%
}
\newcommand*\patchBothAmsMathEnvironmentsForLineno[1]{%
  \patchAmsMathEnvironmentForLineno{#1}%
  \patchAmsMathEnvironmentForLineno{#1*}%
}
\AtBeginDocument{%
\patchBothAmsMathEnvironmentsForLineno{equation}%
\patchBothAmsMathEnvironmentsForLineno{align}%
\patchBothAmsMathEnvironmentsForLineno{flalign}%
\patchBothAmsMathEnvironmentsForLineno{alignat}%
\patchBothAmsMathEnvironmentsForLineno{gather}%
\patchBothAmsMathEnvironmentsForLineno{multline}%
\patchBothAmsMathEnvironmentsForLineno{eqnarray}%
}

\usepackage{hyperref}    
\usepackage[all]{hypcap} 

\def\lhcb {\mbox{LHCb}\xspace}

\def\MagUp {\mbox{\em Mag\kern -0.05em Up}\xspace}

\ifthenelse{\boolean{uprightparticles}}%
{

 \def\PDelta      {\ensuremath{\Delta}\xspace}                 
 \def\PXi      {\ensuremath{\Xi}\xspace}                 
 \def\PLambda      {\ensuremath{\Lambda}\xspace}                 
 \def\PSigma      {\ensuremath{\Sigma}\xspace}                 
 \def\POmega      {\ensuremath{\Omega}\xspace}                 
 \def\PUpsilon      {\ensuremath{\Upsilon}\xspace}                 
 
 \def\PB      {\ensuremath{\mathrm{B}}\xspace}                 
                  
 \def\PD      {\ensuremath{\mathrm{D}}\xspace}

 \def\PK      {\ensuremath{\mathrm{K}}\xspace}

 \def\Pi      {\ensuremath{\mathrm{i}}\xspace}

}
{

 \mathchardef\PDelta="7101
 \mathchardef\PXi="7104
 \mathchardef\PLambda="7103
 \mathchardef\PSigma="7106
 \mathchardef\POmega="710A
 \mathchardef\PUpsilon="7107
                  
 \def\PB      {\ensuremath{B}\xspace}                 
                  
 \def\PD      {\ensuremath{D}\xspace}

 \def\PK      {\ensuremath{K}\xspace}

 \def\Pi      {\ensuremath{i}\xspace}

}

\makeatletter
\ifcase \@ptsize \relax
  \newcommand{\miniscule}{\@setfontsize\miniscule{4}{5}}
\or
  \newcommand{\miniscule}{\@setfontsize\miniscule{5}{6}}
\or
  \newcommand{\miniscule}{\@setfontsize\miniscule{5}{6}}
\fi
\makeatother

\DeclareRobustCommand{\optbar}[1]{\shortstack{{\miniscule (\rule[.5ex]{1.25em}{.18mm})}
  \\ [-.7ex] $#1$}}

\def\antikt     {\ensuremath{\text{anti-}k_{\mathrm{T}}}\xspace}

\def\fastjet    {\mbox{\textsc{FastJet}}\xspace}

  \def\Kbar    {{\kern 0.2em\overline{\kern -0.2em \PK}{}}\xspace}

\def\KorKbar    {\kern 0.18em\optbar{\kern -0.18em K}{}\xspace}

  \def\Dbar    {{\kern 0.2em\overline{\kern -0.2em \PD}{}}\xspace}

\def\DorDbar    {\kern 0.18em\optbar{\kern -0.18em D}{}\xspace}

\def\Bbar    {{\ensuremath{\kern 0.18em\overline{\kern -0.18em \PB}{}}}\xspace}

\def\BorBbar    {\kern 0.18em\optbar{\kern -0.18em B}{}\xspace}

  \def\Y#1S{\ensuremath{\PUpsilon{(#1S)}}\xspace}

\def\Lbar        {{\ensuremath{\kern 0.1em\overline{\kern -0.1em\PLambda}}}\xspace}
\def\LorLbar    {\kern 0.18em\optbar{\kern -0.18em \PLambda}{}\xspace}

\def\to                 {\ensuremath{\rightarrow}\xspace}

\def\AT#1     {\ensuremath{A_{\mathrm{T}}^{#1}}\xspace}           

\def\C#1      {\ensuremath{\mathcal{C}_{#1}}\xspace}                       
\def\Cp#1     {\ensuremath{\mathcal{C}_{#1}^{'}}\xspace}                    
\def\Ceff#1   {\ensuremath{\mathcal{C}_{#1}^{\mathrm{(eff)}}}\xspace}        
\def\Cpeff#1  {\ensuremath{\mathcal{C}_{#1}^{'\mathrm{(eff)}}}\xspace}       
\def\Ope#1    {\ensuremath{\mathcal{O}_{#1}}\xspace}                       
\def\Opep#1   {\ensuremath{\mathcal{O}_{#1}^{'}}\xspace}                    

\newcommand{\tev}{\ifthenelse{\boolean{inbibliography}}{\ensuremath{~T\kern -0.05em eV}\xspace}{\ensuremath{\mathrm{\,Te\kern -0.1em V}}}\xspace}
\newcommand{\gev}{\ensuremath{\mathrm{\,Ge\kern -0.1em V}}\xspace}
\newcommand{\mev}{\ensuremath{\mathrm{\,Me\kern -0.1em V}}\xspace}
\newcommand{\kev}{\ensuremath{\mathrm{\,ke\kern -0.1em V}}\xspace}
\newcommand{\ev}{\ensuremath{\mathrm{\,e\kern -0.1em V}}\xspace}
\newcommand{\gevc}{\ensuremath{{\mathrm{\,Ge\kern -0.1em V\!/}c}}\xspace}
\newcommand{\mevc}{\ensuremath{{\mathrm{\,Me\kern -0.1em V\!/}c}}\xspace}
\newcommand{\gevcc}{\ensuremath{{\mathrm{\,Ge\kern -0.1em V\!/}c^2}}\xspace}
\newcommand{\gevgevcccc}{\ensuremath{{\mathrm{\,Ge\kern -0.1em V^2\!/}c^4}}\xspace}
\newcommand{\mevcc}{\ensuremath{{\mathrm{\,Me\kern -0.1em V\!/}c^2}}\xspace}

\def\mum  {\ensuremath{{\,\upmu\rm m}}\xspace}

\def\gsim{{~\raise.15em\hbox{$>$}\kern-.85em
          \lower.35em\hbox{$\sim$}~}\xspace}
\def\lsim{{~\raise.15em\hbox{$<$}\kern-.85em
          \lower.35em\hbox{$\sim$}~}\xspace}

\def\ptot       {\mbox{$p$}\xspace}
\def\pt         {\mbox{$p_{\rm T}$}\xspace}

\def\evtgen     {\mbox{\textsc{EvtGen}}\xspace}

\def\geant      {\mbox{\textsc{Geant4}}\xspace}

\def\photos     {\mbox{\textsc{Photos}}\xspace}

\def\pythia     {\mbox{\textsc{Pythia}}\xspace}

\def\tell1  {TELL1\xspace}
\def\ukl1   {UKL1\xspace}

\newcommand{\ie}{\mbox{\itshape i.e.}\xspace}

\usepackage{cite} 
\usepackage{mciteplus}

\begin{document}

\renewcommand{\thefootnote}{\fnsymbol{footnote}}
\setcounter{footnote}{1}


\begin{titlepage}
\pagenumbering{roman}

\vspace*{-1.5cm}
\centerline{\large EUROPEAN ORGANIZATION FOR NUCLEAR RESEARCH (CERN)}
\vspace*{1.5cm}
\noindent
\begin{tabular*}{\linewidth}{lc@{\extracolsep{\fill}}r@{\extracolsep{0pt}}}
\ifthenelse{\boolean{pdflatex}}
{\vspace*{-2.7cm}\mbox{\!\!\!\includegraphics[width=.14\textwidth]{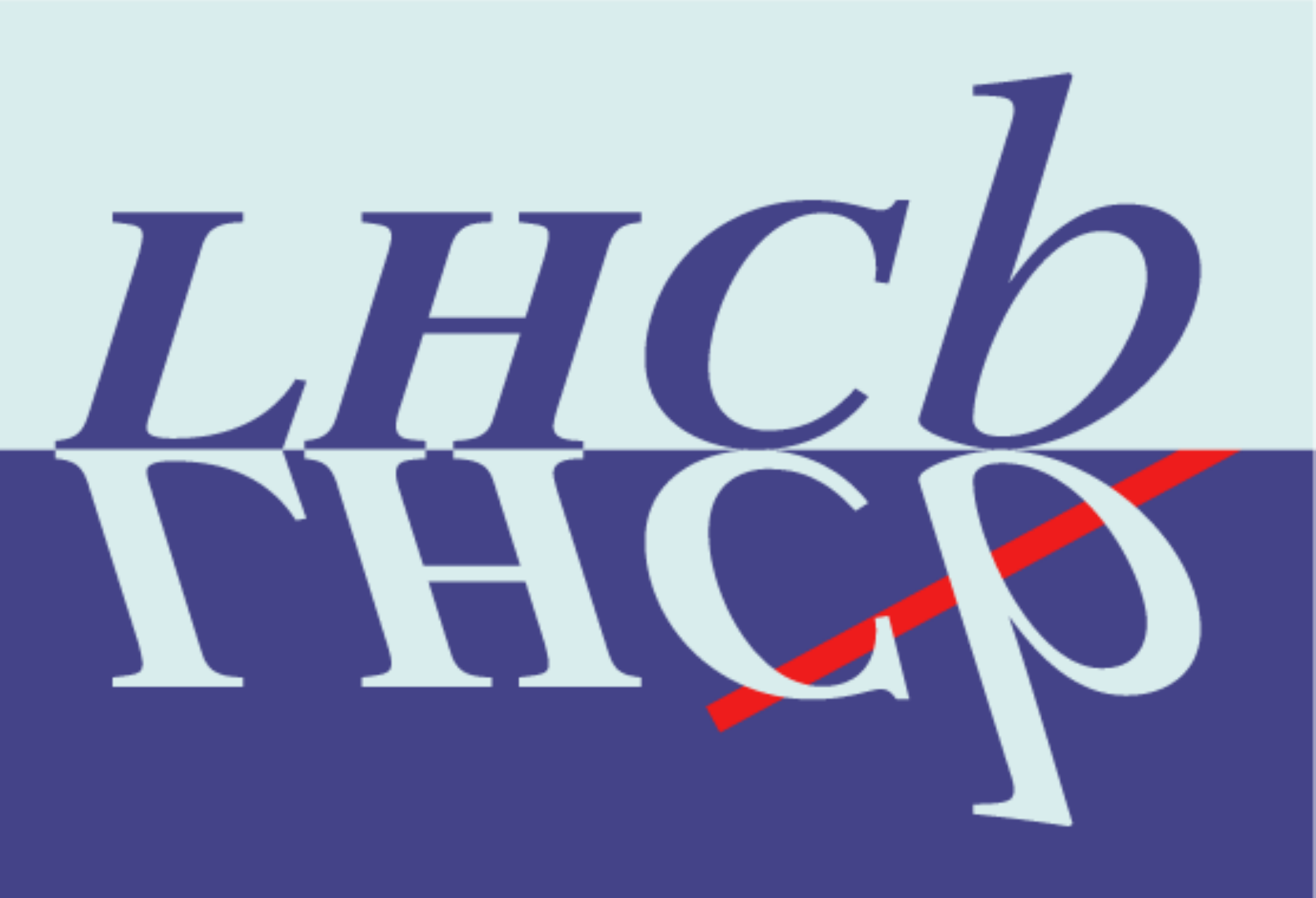}} & &}%
{\vspace*{-1.2cm}\mbox{\!\!\!\includegraphics[width=.12\textwidth]{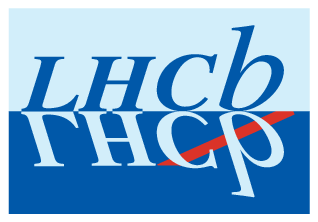}} & &}%
\\
 & & CERN-EP-2017-211 \\  
 & & LHCb-PAPER-2017-024\\  
 & & 5 February 2018 \\ 
 & & \\
\end{tabular*}

\vspace*{4.0cm}

{\normalfont\bfseries\boldmath\huge
\begin{center}
First observation of forward $Z \to b \bar{b}$ production in $pp$ collisions at $\sqrt{s}=8$~TeV 
\end{center}
}

\vspace*{2.0cm}

\begin{center}
The LHCb collaboration\footnote{Authors are listed at the end of this letter.}
\end{center}

\vspace{\fill}

\begin{abstract}
  \noindent
The decay $Z \to b \bar{b}$ is reconstructed in $pp$ collision data, corresponding to 2 fb$^{-1}$ of integrated luminosity, collected by the LHCb experiment at a centre-of-mass energy of $\sqrt{s}=8$ TeV. The product of the $Z$ production cross-section and the $Z \to b \bar{b}$ branching fraction is measured for candidates in the fiducial region defined by two particle-level $b$-quark jets with pseudorapidities in the range $2.2 < \eta < 4.2$, with transverse momenta $\pt>20$ GeV and dijet invariant mass in the range $45 < m_{jj} < 165$ GeV.
From a signal yield of $5462 \pm 763$ $Z \to b \bar{b}$ events, where the uncertainty is statistical, a production cross-section times branching fraction of $332 \pm 46 \pm 59$ pb is obtained, where the first uncertainty is statistical and the second systematic.
The measured significance of the signal yield is 6.0 standard deviations.
This measurement represents the first observation of the $Z \to b \bar{b}$ production in the forward region of $pp$ collisions.
\end{abstract}

\vspace*{2.0cm}

\begin{center}
  Published in Phys. Lett. B776 (2017) 430-439
\end{center}

\vspace{\fill}

{\footnotesize 
\centerline{\copyright~CERN on behalf of the \lhcb collaboration, licence \href{http://creativecommons.org/licenses/by/4.0/}{CC-BY-4.0}.}}
\vspace*{2mm}

\end{titlepage}


\newpage
\setcounter{page}{2}
\mbox{~}
%
%
%
%

\cleardoublepage


\renewcommand{\thefootnote}{\arabic{footnote}}
\setcounter{footnote}{0}



\pagestyle{plain} 
\setcounter{page}{1}
\pagenumbering{arabic}


%

\section{Introduction}
\label{sec:introduction}

Measurements of $Z$-boson production in $pp$ collisions constitute an important test of the Standard Model (SM), since they allow the electroweak sector to be precisely probed \cite{Majorana:1937vz,Pati:1974yy,Mohapatra:1979ia}.
The LHCb experiment can be used to measure the decay of the $Z$ boson into a $b \bar{b}$ quark pair in the forward region that is inaccessible at other LHC experiments.

The decay $Z \to b \bar{b}$ provides a standard candle for searches in final states with a $b \bar{b}$ quark pair. 
The inclusive search for the SM Higgs decay to two $b$ quarks at the LHC is of great interest, since the measurement of the Higgs boson coupling to $b$ quarks is an important test of the SM \cite{TheATLASandCMSCollaborations:2015bln}.
Several extensions of the SM predict that new heavy particles that decay to two energetic $b$ quarks could be accessible at LHC collision energies \cite{Baur:1989kv, Frampton:1987dn, Sirunyan:2017dnz}.
A sizeable $Z \to b \bar{b}$ event sample will enable the measurement of the $b \bar{b}$ forward-central asymmetry at the $Z$ pole, which could be enhanced by the contributions from new physics processes\cite{Grinstein:2013iws}. The forward-central asymmetry in inclusive $b \bar{b}$ events has previously been measured by the LHCb collaboration \cite{Aaij:2014ywa}.

The measurements of this decay can also be used to demonstrate that no biases are induced by the $b$-jet reconstruction procedure and that the reconstruction efficiencies are evaluated correctly. 
In addition, the $Z \to b \bar{b}$ decay is important to determine the so-called $b$-jet energy scale.
This is the factor that has to be applied to the reconstructed $b$-jet energy in simulated events in order to reproduce the actual detector response.

The reconstruction of the $Z \to b \bar{b}$ decay is challenging at hadron colliders, due to the large QCD background.
Many techniques to reconstruct the $Z \to b \bar{b}$ decay channel have been developed by the CDF \cite{Donini:2008nt}, ATLAS \cite{Aad:2014bla} and CMS \cite{CMS:2017cbv} collaborations.
The CDF collaboration reconstructed the $Z \to b \bar{b}$ decay in $p\bar{p}$ collisions at 1.96 TeV and determined the $b$-jet energy scale, obtaining a relative uncertainty on the product of the cross-section and the branching fraction of 29$\% $. 
The analysis of the ATLAS collaboration reconstructed boosted $Z \to b \bar{b}$ candidates in the central region of $pp$ collisions at 8 TeV, with pseudorapidity $|\eta|<2.5$, and determined the cross-section with a relative uncertainty of 16$\%$.
The CMS collaboration made the first observation of the $Z \to b \bar{b}$ decay in a single-jet topology in the same pseudorapidity region, with a significance of 5.1 standard deviations.

This Letter describes a new method to study the $Z\to b\bar{b}$ decay, performed on $pp$ collision data collected at a centre-of-mass energy of $\sqrt{s}=8$ TeV, corresponding to an integrated luminosity of 2 fb$^{-1}$.
The low trigger thresholds on the particle energies that are employed at LHCb and the excellent $b$-jet identification performance make it possible to select candidates within a large invariant mass range, including those with masses below the $Z$-boson pole.
Events are selected requiring two $b$-jet candidates, referred to as a $b$ dijet, and an additional jet that balances the transverse momentum of the $b \bar{b}$ system. The invariant mass distribution of the $b$ dijet is used to determine the $Z \to b \bar{b}$ yield and the $b$-jet energy scale. The invariant mass distribution of the QCD background is determined using a control region that is defined through observables related to the $b$-dijet system and to the associated balancing jet. Simulated data are used to evaluate the reconstruction efficiency and the detector acceptance, enabling a measurement of the $Z$  production cross-section multiplied by the $Z \to b \bar{b}$ branching fraction.

\section{The LHCb detector, trigger and simulation}
\label{sec:detector}
The \lhcb detector~\cite{Alves:2008zz,LHCb-DP-2014-002} is a single-arm forward spectrometer fully instrumented in the \mbox{pseudorapidity} range $2<\eta <5$, which is designed for the study of $b$ and $c$ hadrons. The detector includes a high-precision tracking system consisting of a silicon-strip vertex detector surrounding the $pp$ interaction region, a silicon-strip detector located upstream of a dipole magnet with a bending power of about $4{\mathrm{\,Tm}}$, and three stations of silicon-strip detectors and straw drift tubes placed downstream of the magnet. The tracking system provides a measurement of momentum, \ptot, of charged particles with a relative uncertainty that varies from 0.5\% at low momentum to 1.0\% at 200\gev.\footnote{In this Letter natural units where $\hbar$ = $c$ = 1 are used.} 
The minimum distance of a track to a primary vertex, the impact parameter, is measured with a resolution of $(15+29/\pt)\mum$, where \pt is the component of the momentum transverse to the beam, in\,\gev.
Different types of charged hadrons are distinguished using information from two ring-imaging Cherenkov detectors. Photons, electrons and hadrons are identified by a calorimeter system consisting of scintillating-pad (SPD) and preshower detectors, an electromagnetic calorimeter and a hadronic calorimeter. Muons are identified by a system composed of alternating layers of iron and multiwire proportional chambers. The online event selection is performed by a trigger system, which consists of a hardware stage, based on information from the calorimeter and muon systems, followed by a software stage, which applies a full event reconstruction.
 
Events are required to satisfy at least one of the following hardware trigger requirements: contain a muon with $\pt>1.86\gev$, a hadron with transverse energy in the calorimeters $E_\mathrm{T} >3.7\gev$, an electron with $E_\mathrm{T}>3\gev$, a photon with $E_\mathrm{T}>3\gev$ or a pair of muons with $\pt_1 \cdot \pt_2 >1.6\gev^2$.
A global event cut~(GEC) on the number of hits in the SPD is applied in order to prevent high-multiplicity events from dominating the processing time.
At the software trigger stage events are required to have a two-, three- or four-track secondary vertex (SV) with significant displacement from any primary vertex. A multivariate algorithm \cite{Likhomanenko:2015aba} is used for the identification of secondary vertices consistent with the decay of a $b$ hadron, strongly suppressing the contamination from charmed hadrons.    

Simulated events generated with \pythia ~\cite{Sjostrand:2007gs,*Sjostrand:2006za}, with a specific \lhcb configuration~\cite{LHCb-PROC-2010-056}, are used to model the properties of the signal $Z \to b \bar{b}$ events and backgrounds such as $Z \to c \bar{c}$, $W \to q q'$ decays and $t\bar{t}$ events.
Decays of hadronic particles are described by \evtgen~\cite{Lange:2001uf}, where the final-state radiation is generated using \photos~\cite{Golonka:2005pn}.
The interaction of the generated particles with the detector, and its response, are implemented using the \geant toolkit~\cite{Allison:2006ve, *Agostinelli:2002hh} as described in Ref.~\cite{LHCb-PROC-2011-006}.

\section{Candidate selection}
\label{sec:sel}

Candidates are selected by requiring the presence of at least three jets, which are reconstructed as detailed in Refs.~\cite{LHCb-PAPER-2015-022,LHCb-PAPER-2015-021,LHCb-PAPER-2016-011,LHCb-PAPER-2013-058}.
Jets are reconstructed using a particle flow algorithm~\cite{LHCb-PAPER-2013-058} and are clustered with the \antikt algorithm~\cite{antikt} with a distance parameter~$0.5$, as implemented in the \fastjet software package~\cite{fastjet}. A jet energy correction~\cite{LHCb-PAPER-2013-058} determined from simulation is applied to recover the jet energy at particle level and jet quality requirements are applied~\cite{LHCb-PAPER-2013-058}. Jets are heavy-flavour tagged, \ie as containing a $b$ or $c$ hadron, if a SV is found with a distance $\Delta R<0.5$ from the jet axis, where $\Delta R$ is the distance in the ($\eta,\phi$) plane and $\phi$ is the azimuthal angle between the jet axis and the vector that points from the $pp$ interaction point to the SV. The details of the flavour-tagging algorithm are described in Ref.~\cite{LHCb-PAPER-2015-016}.
Two heavy-flavour tagged jets are required to form a $Z \to b \bar{b}$ candidate.
At least one of the two $b$-jet candidates must be tagged by a SV selected by the software trigger requirements.
The two heavy-flavour jets are each required to have transverse momenta $\pt>20\gev$, pseudorapidities in the range $2.2<\eta<4.2$, and a combined invariant mass ($m_{jj}$) in the range $45 < m_{jj} < 165$~GeV.
The fiducial region of the measurement within which the cross-section is determined is defined by the kinematical requirements described above applied to particle-level jets, which are jets reconstructed in the simulation from stable particles (\emph{i.e.} particles with lifetime in excess of 10 ps, excluding neutrinos) using the default reconstruction algorithm.

In order to increase the signal-to-background ratio, the absolute azimuthal angle between the two $b$-jets is required to be greater than 2.5 radians.
The presence of a balancing jet is required to help discriminate $Z \to b \bar{b}$ events from the QCD multijet background.
The $Z$ + jet signal is predominantly produced via quark-gluon scattering, while the QCD multijet background is produced via gluon-gluon interactions~\cite{Campbell:2006wx}. 
The balancing jet is defined as that which minimises the total $\pt$ of the $Z$ boson and the jet.
This jet is required to have $\pt>10\gev$ and $2.2<\eta<4.2$.
Given the SM cross-sections \cite{PDG2014} and the selection efficiencies, which are evaluated using simulation, about 17$\times 10^3$ $Z \to b \bar{b}$ candidates, 600 $Z \to c \bar{c}$ candidates, 200 $W \to q q'$ candidates and 50 $t \bar{t}$ candidates are expected after the application of the selection criteria. 
A sample of around $6 \times 10^5$ candidates is selected in data, dominated by the combinatorial background from the multijet QCD events.

A multivariate classifier is trained to discriminate $Z \to b \bar{b}$ events from combinatorial QCD events.
A uniform Gradient Boost Boosted Decision Tree technique~\cite{Rogozhnikov:2014zea} is adopted, in order to ensure a selection efficiency with a low dependence on the dijet invariant mass. 
The classifier is trained using four kinematical variables of the three-jet system, chosen for both their low correlation with the dijet invariant mass and for their discriminating power.
The variables are the absolute pseudorapidity difference between the two heavy-flavour jets, the $\pt$ of the balancing jet, the angle between the balancing-jet momentum and the $Z$-boson candidate momentum in the azimuthal plane with respect to the beam axis, and the polar angle between the balancing-jet momentum and the $Z$-boson flight direction in the $Z$-boson rest frame.
The classifier is trained using 5$\%$ of the data sample to represent the combinatorial QCD background. 
This training sample has a negligible $Z \to b \bar{b}$, $Z \to c \bar{c}$, $W \to q q'$ and $t \bar{t}$ contamination and it is not used in the dijet invariant mass fit described below.
The signal process is modelled using simulated $Z \to b \bar{b}$ events.
The distributions of the input observables related to the balancing-jet kinematics are validated by comparing the high purity $Z(\to \mu^+ \mu^-)$ + jet data sample described in Ref.~\cite{LHCb-PAPER-2016-011} with the corresponding simulation sample.

\begin{figure}[h]
\begin{center}
\includegraphics[width=10cm]{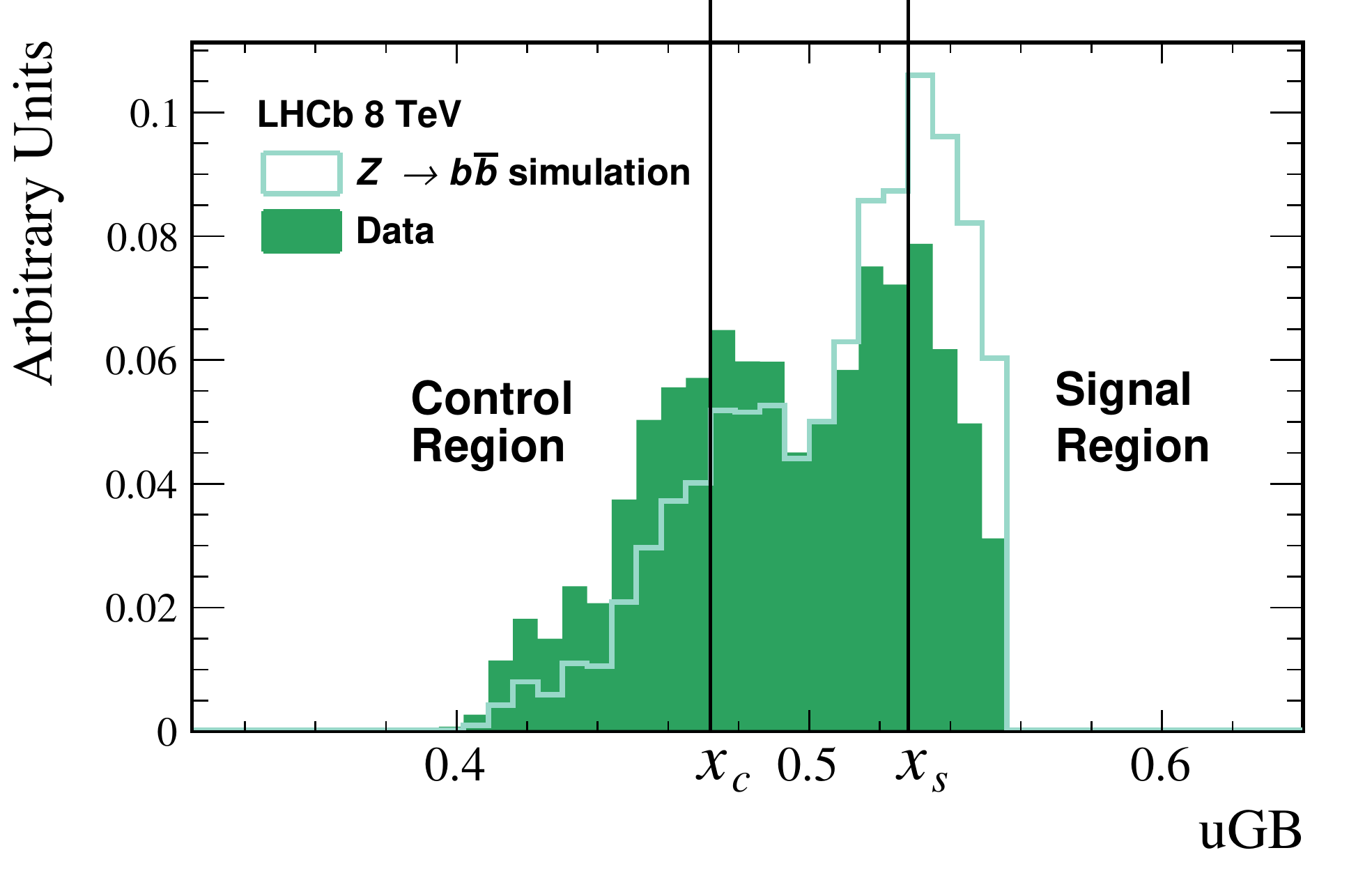}
\caption{\label{fig:ugb} Distribution of the multivariate classifier output for data and for simulated $Z \to b \bar{b}$ decays, normalisted to unity. The signal region is defined by $\mathrm{uGB}>x_s$ and the control region by events with $\mathrm{uGB}<x_c$.}
\end{center}
\end{figure}

The output of the classifier ($\mathrm{uGB}$) is shown in Fig.~\ref{fig:ugb}. 
Candidates are selected in two different regions of uGB: the signal region ($\mathrm{uGB}>x_s$), which has enhanced $Z \to b \bar{b}$ contribution, and a control region ($\mathrm{uGB}<x_c$), which has a larger contribution from QCD combinatorial events. The two regions are fitted simultaneously to determine the $Z \to b \bar{b}$ yield, and the values of $x_s$ and $x_c$ are chosen in order to achieve the best signal significance.

\section{Signal yield determination}
\label{sec:fit}

A simultaneous fit to the $b$-dijet invariant mass distributions in the signal and control regions is performed to determine the $Z \to b \bar{b}$ yield and the jet energy scale factor, $k_\mathrm{JES}$. A triple-Gaussian model is used to describe the $Z \to b \bar{b}$ dijet invariant mass distribution. The parameters of this model are obtained separately for the candidates in the signal and control regions using simulation, and are fixed in the fit to the data.
The $k_\mathrm{JES}$ factor is also introduced in the $Z \to b \bar{b}$ invariant mass distribution model in order to account for differences between simulation and data in the jet four-momentum. 
This is achieved by substituting $m_{jj}$ with $m_{jj}/k_{\mathrm{JES}}$ in the model.
The reconstructed invariant mass of dijets in $Z \to b \bar{b}$ simulated events has a mean of 80 GeV, \emph{i.e.} below the known $Z$-boson mass \cite{PDG2014}, and a resolution of 16$\%$. 
The reduced mean is due to parton radiation outside the jet cone, missing energy, and residual biases in the reconstructed jet energy that are not recovered by the jet energy correction. 

The invariant mass distribution of the combinatorial background is parametrized with a Pearson IV distribution, as is typical to describe the multijet combinatorial background \cite{Donini:2008nt}.
The four parameters of the Pearson IV function are free to vary in the fit and they have approximately the same values in the signal and control regions, since the uGB is trained to be as  uniform as possible with respect to the dijet invariant mass. To take into account the residual correlation with the dijet invariant mass, the Pearson IV distribution is multiplied in the signal (control) region by a linear transfer function $t^{s(c)}(m_{jj})$, defined as
\begin{displaymath}
t^{s(c)}(m_{jj}) = a^{s(c)}+b^{s(c)} \cdot m_{jj},
\end{displaymath}
where the superscript $s$ ($c$) indicates the signal (control) region, and $a^{s(c)}$ and $b^{s(c)}$ are parameters fixed in the invariant mass fit.
The parameters $a^{s(c)}$ and $b^{s(c)}$ are determined by fitting the transfer function to the selection efficiency after the requirement that $\mathrm{uGB}>x_s$ ($\mathrm{uGB}<x_c$) as a function of the dijet invariant mass in the $45<m_{jj}<60$ GeV and $100<m_{jj}<165$ GeV intervals, where the $Z \to b \bar{b}$ contribution is negligible. 
As a cross-check, data events with $\mathrm{uGB} < x_c$ are fitted with only the QCD background model, ignoring the small $Z \to b \bar{b}$ contribution, and a good fit quality is obtained.

The invariant mass model used to fit the signal region is
\begin{displaymath}
f^s(m_{jj}) = N_Q^s Q(m_{jj}) \cdot t^s(m_{jj}) + N_Z^s Z^s(m_{jj}; k_\mathrm{JES}),
\end{displaymath}
where $N_Q^s$ and $N_Z^s$ are the number of QCD events and the number of $Z$-boson events ($Z \to b \bar{b}$ plus $Z \to c \bar{c}$) in the signal region respectively, and $Q(m_{jj})$, $t^s(m_{jj})$ and $Z^s(m_{jj})$ are the Pearson IV distribution, the transfer function and the $Z$-boson invariant mass distribution model in the signal region, respectively. The $Z \to c \bar{c}$ invariant mass distribution is assumed to be identical to that of $Z \to b \bar{b}$ events. This assumption is verified using the simulation and the two components are therefore fitted together. Backgrounds other than $Z \to c \bar{c}$ and QCD multijet events are neglected in the fit. 
Since the $\mathrm{uGB} > x_s$ requirement is applied, the expected value of $N_Z^s$ is lower than the 17$\times 10^3$ $Z \to b \bar{b}$ events expected before the uGB selection.

The invariant mass model that describes the control region is
\begin{displaymath}
f^c(m_{jj}) = N_Q^c Q(m_{jj}) \cdot t^c(m_{jj}) + R \cdot N_Z^s Z^c(m_{jj}; k_\mathrm{JES}),
\end{displaymath}
where $N_Q^c$ is the number of QCD events in the control region and $Q(m_{jj})$, $t^c(m_{jj})$ and $Z^c(m_{jj})$ are the Pearson IV distribution, the transfer function and the $Z$-boson invariant mass distribution model in the control region. The parameter $R$ is the ratio of the efficiency for $Z$-boson candidates selected with $\mathrm{uGB} < x_c$ and $\mathrm{uGB} > x_s$ and is determined from simulation and fixed in the fit.
A simultaneous unbinned maximum likelihood fit is performed with the $N_Q^s$, $N_Q^c$, $N_Z^s$, $k_\mathrm{JES}$ and the Pearson IV parameters free to vary. 
Pseudoexperiments are used to verify that the fit is stable and estimate any bias. The parameter $N_Z^s$ is determined with a bias of about $2\%$ and the value returned by the fit is corrected accordingly in the cross-section determination. 

\begin{figure}[!t]
\begin{center}
\includegraphics[width=\textwidth]{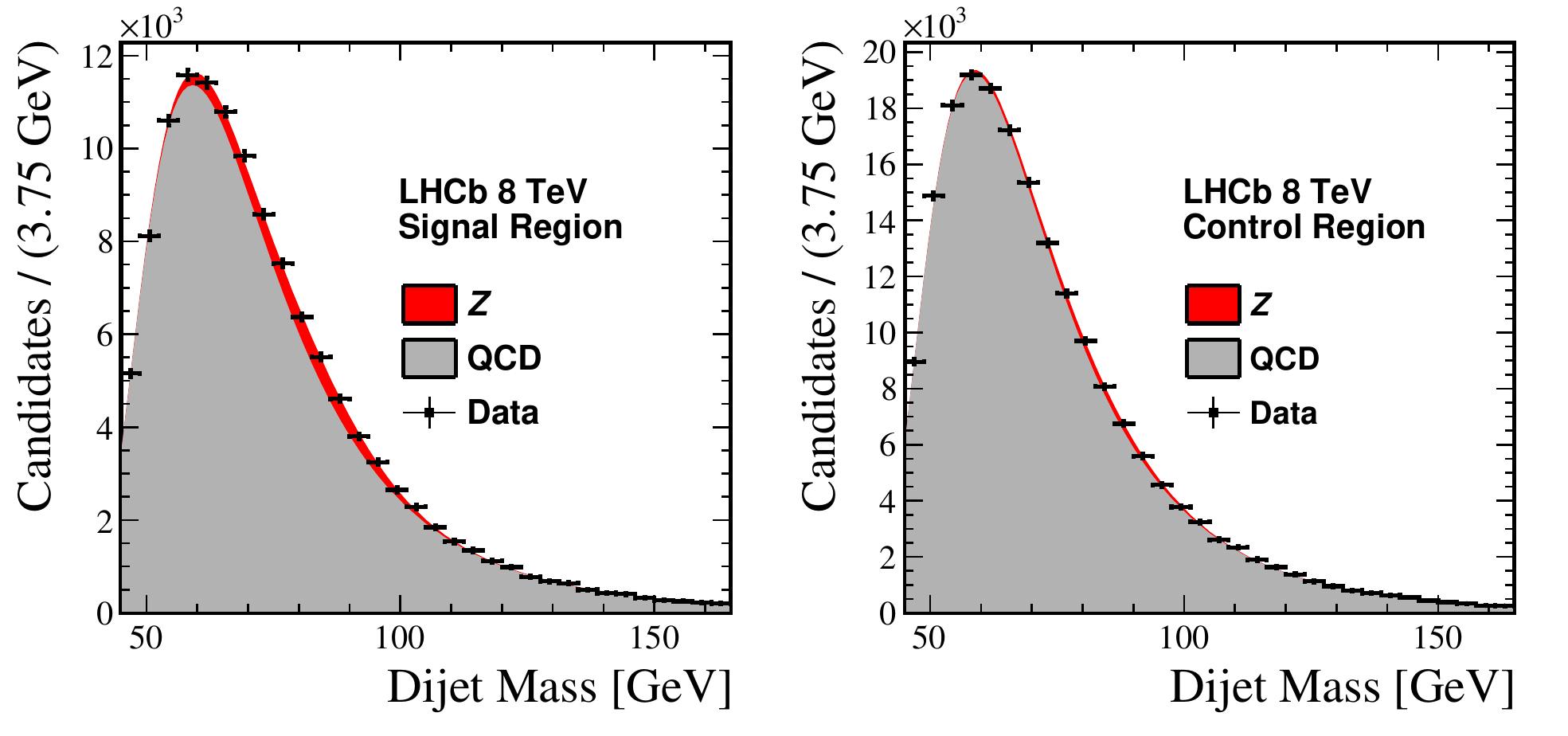}
\caption{\label{fig:fit_result} Simultaneous fit to the dijet invariant mass distribution of $Z \to b \bar{b}$ candidates in the (left) signal and (right) control regions.}
\end{center}
\end{figure}

\begin{figure}[!t]
\begin{center}
\includegraphics[width=\textwidth]{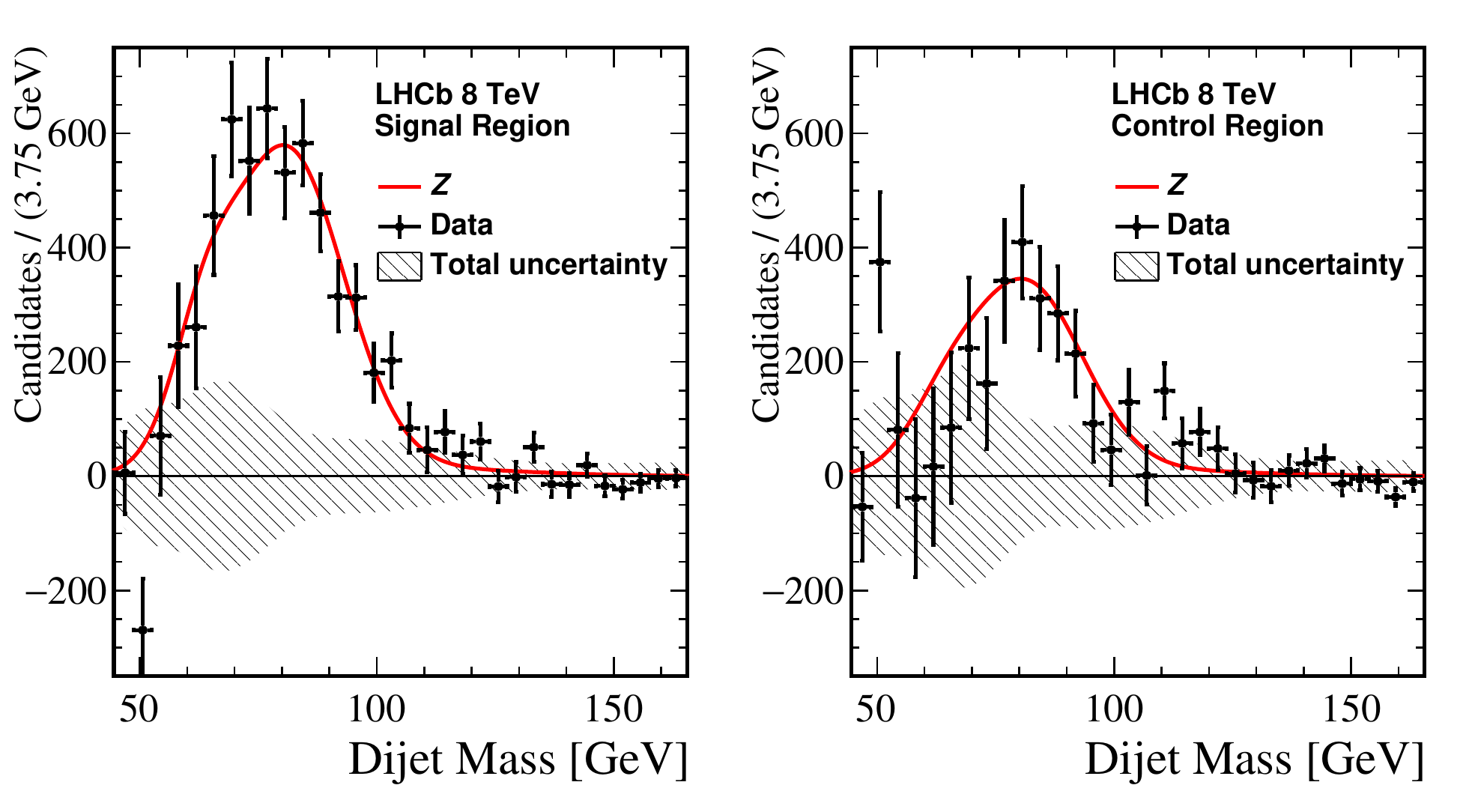}
\caption{\label{fig:fit_peak} Background-subtracted distribution compared with the $Z \to b \bar{b}$ mass model in the (left) signal and (right) control regions. The one standard deviation total uncertainty band in the background-only hypothesis is also shown. This band includes statistical and systematic uncertainties.}
\end{center}
\end{figure}

The fit result is shown in Fig.~\ref{fig:fit_result} and the background-subtracted data and result of the fit are shown in Fig.~\ref{fig:fit_peak}.
The $Z$-boson yield in the signal region is 5462~$\pm$ 763 and the jet energy scale factor is measured to be 1.009 $\pm$ 0.015.
Using Wilks' theorem \cite{Wilks}, the $Z \to b \bar{b}$ statistical significance is found to be 7.3 standard deviations.

As an additional cross-check to validate the technique, a fit to the dijet invariant mass distribution for candidates with $x_c < \mathrm{uGB} < x_s$ is performed, with a model analogous to that used in the signal and control regions. In this case, the parameters of the QCD background are fixed to the values returned by the default fit, but the $Z \to b \bar{b}$ yield in this region, $N_Z^v$, is left free. The goodness of this fit is acceptable and the ratio $N_Z^v/N_Z^s$ is compatible with the expectation from simulation.

\section{Cross-section determination and systematic uncertainties}
\label{systematics}
The product of the $Z$-boson production cross-section and the $Z \to b \bar{b}$ branching fraction is determined using
\begin{displaymath}
\sigma(pp \to Z) \mathcal{B}(Z \to b \bar{b}) = \frac{N_Z^s }{\mathcal{L} \cdot (1-f_{\mathrm{uGB}}) \cdot \epsilon_Z^s  \cdot (1+f_{Z \to c \bar{c}}) } 
\end{displaymath} 
where $\mathcal{L}$ is the integrated luminosity, $\epsilon_Z^s$ is the efficiency of the selection requirements, including $\mathrm{uGB}>x_s$, for events in the fiducial region, $f_{\mathrm{uGB}}$ is the fraction (5\%) of data events removed for the multivariate classifier training and 1+$f_{Z \to c \bar{c}}$ is a factor applied to correct for the small $Z \to c \bar{c}$ contamination. 
The selection efficiency is obtained from simulation, but correction factors are applied to account for differences in the heavy-flavour tagging efficiencies between data and simulation~\cite{LHCb-PAPER-2015-016}. 
By using a small sample with a looser trigger requirement and a technique similar to that described in Ref.~\cite{LHCb-PAPER-2016-011}, the GEC efficiency is also corrected for differences in data and simulation.  
The balancing-jet selection efficiency is corrected at Next-to-Leading-Order (NLO) using simulated $Z \to b \bar{b}$ events produced with aMC@NLO \cite{Alwall:2014hca} plus \pythia for parton showers.
The $f_{Z \to c \bar{c}}$ fraction is obtained by multiplying the $Z \to c \bar{c}$ and $Z \to b \bar{b}$ branching fraction ratio \cite{PDG2014} by the acceptance and the efficiency ratios, both determined using simulation.

The sources of systematic uncertainty considered for the measurement are given in Table~\ref{tab:systematics}.
\begin{table}[t!]
  \begin{center}
      \caption{\label{tab:systematics} Systematic uncertainties on the cross-section, $\sigma_Z=\sigma(pp \to Z) \mathcal{B}(Z \to b \bar{b})$, and jet energy scale in percent. The total uncertainty is the sum in quadrature of all the contributions.}
      \begin{tabular}{c|c|c}
        \toprule
        Systematic source & $\sigma_{Z}$ [$\%$]& $k_\mathrm{JES}$ [$\%$] \\
        \midrule
        Heavy-flavour tagging efficiency & $16.6$ &  ~$\phantom{<}0.5$  \\
        Hardware trigger efficiency & $\phantom{1}1.9$ & ~~~-- \\
        GEC efficiency & $\phantom{1}1.7$ & ~~~-- \\
        Jet energy correction & $\phantom{1}2.7$ & ~$\phantom{<}0.3$ \\
        Jet energy resolution & $\phantom{1}1.0$ & ~$\phantom{<}0.2$  \\
        Jet identification efficiency & $\phantom{1}2.0$ & $<0.1$  \\
        Balancing-jet selection efficiency & $\phantom{1}1.8$ &  ~~~-- \\
        Signal model & $\phantom{1}2.0$ & ~$\phantom{<}0.3$  \\
        QCD model & $\phantom{1}1.1$ & $<0.1$  \\     
        Transfer functions & $\phantom{1}1.5$ & ~$\phantom{<}0.8$ \\        
        $R$ efficiencies ratio & $\phantom{1}0.3$ & $<0.1$ \\
        Fit bias & $\phantom{1}2.1$ & ~~~-- \\
        Subdominant backgrounds ($t \bar{t}$, $W \to q q'$) & $\phantom{1}1.9$ & $<0.1$ \\     
        Final-state radiation & $\phantom{1}0.9$ &  ~~~-- \\               
        $f_{Z \to c \bar{c}}$ fraction & $\phantom{1}0.1$ &  ~~~-- \\        
        Luminosity & $\phantom{1}1.2$ &  ~~~-- \\        
        \midrule
        Total & $17.7$ & $~\phantom{<}1.1$  \\ 
        \bottomrule
      \end{tabular}
  \end{center}
\end{table}
Systematic effects that are associated with differences between data and simulation can affect the signal invariant mass distribution model and the selection efficiency.
The impact of these differences is evaluated by repeating the fit with a modified signal model and by recalculating the cross-section varying $\epsilon_Z^s$.
Other sources of systematic uncertainties are related to the signal extraction procedure. 

The method described in Ref.~\cite{LHCb-PAPER-2015-016} is used to assess the systematic uncertainty due to the heavy-flavour tagging efficiency which amounts to $5\%-10\%$ per jet, depending on the $\pt$ range. 
This uncertainty is dominated by the size of the calibration samples used in the heavy-flavour tagging efficiency measurement.
Since one of the two $b$-jet candidates must be tagged by a SV selected by the software trigger, the uncertainty on this trigger efficiency is included in this contribution.
The systematic uncertainty associated with the hardware trigger efficiency is determined by measuring the efficiency with a tag-and-probe technique, using the high purity $Z(\to \mu^+ \mu^-)$ + jet data sample~\cite{LHCb-PAPER-2016-011}. In order to avoid trigger bias on the jet selection, the tag is the muon that triggered the event and the probe is the associated jet. The hardware trigger efficiency measured on probe jets is compared between data and simulation and the maximum difference in intervals of the jet \pt is taken as an uncertainty.
The latter does not take into account the systematic uncertainty on the GEC efficiency, which is determined separately by studying its dependence on the $b$-dijet invariant mass and assigning the largest variation as the uncertainty.
The systematic uncertainty on the jet energy correction includes biases due to jet flavour dependence, reconstruction of tracks which are not associated to a real particle, the track momentum resolution and residual differences between simulation and data, as described in Refs.~\cite{LHCb-PAPER-2016-011,LHCb-PAPER-2013-058}. The jet energy resolution is modelled in simulation with an uncertainty measured in Refs.~\cite{LHCb-PAPER-2015-021,LHCb-PAPER-2013-058}. The uncertainties related to the jet reconstruction and identification are taken from Ref.~\cite{LHCb-PAPER-2013-058}.
The systematic uncertainty associated with the balancing-jet selection efficiency is evaluated by measuring this efficiency in the $Z(\to \mu^+ \mu^-)$ + jet data and simulation samples and taking the difference as a systematic uncertainty. 

The uncertainty on the model of the signal invariant mass distribution is determined by repeating the fit with an alternative distribution, consisting of the sum of two modified Gaussians.
The uncertainty on the QCD model is determined by considering an alternative parametrization, consisting of an exponential decay model multiplied by a function that describes the effect of the jet \pt requirements on the invariant mass distribution. It has been verified, by generating pseudoexperiments with this alternative model and by fitting them with the default model, that the choice of the QCD distribution model introduces a small bias in the measurement. This bias is taken as the systematic uncertainty.
The systematic uncertainty associated with the transfer functions is evaluated by repeating the fit using second-order polynomial functions instead of linear functions. In these fits the coefficients of the quadratic terms are varied in a range consistent with the data in the invariant mass sidebands used in the determination of the transfer functions. The maximum variation with respect to the default measurement is taken as the uncertainty.
The efficiency ratio $R$ is determined using both $Z(\to \mu^+ \mu^-)$ + jet data and simulation, and the observed difference is taken as a systematic uncertainty.
The uncertainty associated with a possible bias introduced by the fit procedure is determined using pseudoexperiments.

The fit is repeated introducing contributions from the subdominant backgrounds, $t \bar{t}$ and $W \to q q'$, fixed to their SM expectations \cite{PDG2014} and modelled with the simulation. 
The difference in the results is assigned as a systematic uncertainty.
The final-state radiation systematic uncertainty is determined as described in Ref.~\cite{Sjostrand:2007gs,*Sjostrand:2006za}.
The systematic uncertainty due to the $Z \to c \bar{c}$ contribution is dominated by the knowledge of the $Z \to c \bar{c}$  branching fraction~\cite{PDG2014} used in the evaluation of the $f_{Z \to c \bar{c}}$ parameter.
The systematic uncertainty on the luminosity is determined as in Ref.~\cite{LHCb-PAPER-2014-047}.

The different sources of systematic uncertainties are considered to be uncorrelated and the total, relative systematic uncertainty is $17.7\%$ for the cross-section measurement, dominated by the heavy-flavour tagging efficiency uncertainty ($16.6\%$). The total systematic uncertainty for the jet energy scale measurement is $1.1\%$ and is dominated by the uncertainty on the transfer functions ($0.8\%$). 
The significance of the signal yield, including all statistical and systematic uncertainties, is 6.0 standard deviations.

\section{Results and conclusions}
\label{sec:conclusions}
The product of the $Z$-boson production cross-section and the $Z \to b \bar{b}$ branching fraction in $pp$ collisions at a centre-of-mass energy of 8 $\mathrm{TeV}$ is 
\begin{displaymath}
\sigma(pp \to Z) \mathcal{B}(Z \to b \bar{b}) = 332 \pm 46 \pm 59  ~ \mathrm{pb},
\end{displaymath}
where the first uncertainty is statistical and the second is systematic.
The measurement is made in the fiducial region defined by two particle-level $b$ jets with $\pt>20\gev$, $2.2<\eta<4.2$, and $45 < m_{jj} < 165$~GeV.

The expected cross-section in the fiducial region of the experimental measurement is calculated at NLO using aMC@NLO plus \pythia for the parton showers and the NNPDF3.0 Parton Distribution Functions (PDFs) set \cite{Ball:2014uwa}. The theoretical prediction determined in this way is
\begin{displaymath}
\sigma(pp \to Z) \mathcal{B}(Z \to b \bar{b}) = 272  ^{+9}_{-12}(\mathrm{scale}) \pm 5 (\mathrm{PDFs}) ~ \mathrm{pb},
\end{displaymath}
where the first uncertainty is related to the missing higher-order corrections and to the value of the strong coupling constant, and the second uncertainty is related to the PDFs.
The uncertainty due to missing higher-order corrections is evaluated by varying the renormalization and factorization scales by a factor of two around the nominal choice, and taking the maximum differences with respect to the nominal values. The uncertainty on the strong coupling is included by varying it within its uncertainty and recalculating the cross-section.
The uncertainty on the PDFs is estimated by taking the variance of the cross-section predictions, where each replica of the NNPDF3.0 set is used in turn.
The prediction and the measurement are compatible within one standard deviation.
The additional data being collected by the LHCb collaboration will allow a more stringent comparison with the theoretical prediction in the future. Moreover, the systematic uncertainty on the heavy-flavour tagging efficiency will be reduced by collecting more data~\cite{LHCb-PAPER-2015-016}.

The measured jet energy scale factor is
\begin{displaymath}
k_{\mathrm{JES}} = 1.009 \pm 0.015 \pm 0.011,
\end{displaymath} 
where the first uncertainty is statistical and the second uncertainty is systematic.
The $k_\mathrm{JES}$ factor is compatible with unity, which demonstrates that the LHCb simulation reproduces accurately the $b$-jet energy in data for $b \bar b$-jet pairs with about 100 GeV of invariant mass. Since a jet energy correction evaluated using simulation is already applied on $b$ jets, $k_\mathrm{JES}$ represents the residual correction obtained using the data.

\FloatBarrier

\section*{Acknowledgements}
%
%
\noindent We express our gratitude to our colleagues in the CERN
accelerator departments for the excellent performance of the LHC. We
thank the technical and administrative staff at the LHCb
institutes. We acknowledge support from CERN and from the national
agencies: CAPES, CNPq, FAPERJ and FINEP (Brazil); MOST and NSFC
(China); CNRS/IN2P3 (France); BMBF, DFG and MPG (Germany); INFN
(Italy); NWO (The Netherlands); MNiSW and NCN (Poland); MEN/IFA
(Romania); MinES and FASO (Russia); MinECo (Spain); SNSF and SER
(Switzerland); NASU (Ukraine); STFC (United Kingdom); NSF (USA).  We
acknowledge the computing resources that are provided by CERN, IN2P3
(France), KIT and DESY (Germany), INFN (Italy), SURF (The
Netherlands), PIC (Spain), GridPP (United Kingdom), RRCKI and Yandex
LLC (Russia), CSCS (Switzerland), IFIN-HH (Romania), CBPF (Brazil),
PL-GRID (Poland) and OSC (USA). We are indebted to the communities
behind the multiple open-source software packages on which we depend.
Individual groups or members have received support from AvH Foundation
(Germany), EPLANET, Marie Sk\l{}odowska-Curie Actions and ERC
(European Union), ANR, Labex P2IO, ENIGMASS and OCEVU, and R\'{e}gion
Auvergne-Rh\^{o}ne-Alpes (France), RFBR and Yandex LLC (Russia), GVA,
XuntaGal and GENCAT (Spain), Herchel Smith Fund, the Royal Society,
the English-Speaking Union and the Leverhulme Trust (United Kingdom).

\addcontentsline{toc}{section}{References}
\setboolean{inbibliography}{true}
\bibliographystyle{LHCb}
\bibliography{main,LHCb-PAPER}

\newpage


\centerline{\large\bf LHCb collaboration}
\begin{flushleft}
\small
R.~Aaij$^{40}$,
B.~Adeva$^{39}$,
M.~Adinolfi$^{48}$,
Z.~Ajaltouni$^{5}$,
S.~Akar$^{59}$,
J.~Albrecht$^{10}$,
F.~Alessio$^{40}$,
M.~Alexander$^{53}$,
A.~Alfonso~Albero$^{38}$,
S.~Ali$^{43}$
% ...

{\footnotesize \it
% ...
$ ^{5}$Clermont Universit{\'e}, Universit{\'e} Blaise Pascal, CNRS/IN2P3, LPC, Clermont-Ferrand, France\\
% ...
$ ^{10}$Fakult{\"a}t Physik, Technische Universit{\"a}t Dortmund, Dortmund, Germany\\
% ...
$ ^{38}$ICCUB, Universitat de Barcelona, Barcelona, Spain\\
$ ^{39}$Universidad de Santiago de Compostela, Santiago de Compostela, Spain\\
$ ^{40}$European Organization for Nuclear Research (CERN), Geneva, Switzerland\\
% ...
$ ^{43}$Nikhef National Institute for Subatomic Physics, Amsterdam, The Netherlands\\
% ...
$ ^{48}$H.H. Wills Physics Laboratory, University of Bristol, Bristol, United Kingdom\\
% ...
$ ^{53}$School of Physics and Astronomy, University of Glasgow, Glasgow, United Kingdom\\
% ...
$ ^{59}$University of Cincinnati, Cincinnati, OH, United States\\
% ...
}
\end{flushleft}

\end{document}